\documentstyle[aps,epsf]{revtex}

\def\superk{Super--Kamiokande}
\def\cherenkov{Cherenkov}
\newcommand{\cllimitny}[3] {\(\tau/B_{#1} > #2 \times 10^{#3}\)}
\newcommand{\cllimit}[3] {\cllimitny{#1}{#2}{#3} years (90\% CL)}
\def\Mpeppo{p \rightarrow e^+ \pi^0}
\def\Mpmppo{p \rightarrow \mu^+ \pi^0}
\def\Mpepeta{p \rightarrow e^+ \eta}
\def\Mpmpeta{p \rightarrow \mu^+ \eta}
\def\Mnnueta{n \rightarrow \bar{\nu} \eta}
\def\Mpnukp{p \rightarrow \bar{\nu} K^+}
\def\Mpmpko{p \rightarrow \mu^+ K^0}
\def\Mpepko{p \rightarrow e^+ K^0}

\def\peppo{\(\Mpeppo{}\)}
\def\pmppo{\(\Mpmppo{}\)}
\def\pepeta{\(\Mpepeta{}\)}
\def\pmpeta{\(\Mpmpeta{}\)}
\def\nnueta{\(\Mnnueta{}\)}
\def\pnukp{\(\Mpnukp{}\)}
\def\pmpko{\(\Mpmpko{}\)}
\def\pepko{\(\Mpepko{}\)}

\def\Mnumunog{(\not\!\gamma)K^+ \rightarrow (\not\!\gamma)\mu^+ \nu_{\mu}}
\def\Mpnukpnumunog{\Mpnukp ;\; \Mnumunog}
\def\Mnumug{(\gamma)K^+ \rightarrow (\gamma)\mu^+ \nu_{\mu}}
\def\Mpnukpnumug{\Mpnukp ;\; \Mnumug}
\def\Mnumu{K^+ \rightarrow \mu^+ \nu_{\mu}}
\def\Mpnukpnumu{\Mpnukp ;\; \Mnumu}
\def\Mpippio{K^+ \rightarrow \pi^+ \pi^0}
\def\Mpnukppippio{\Mpnukp ;\; \Mpippio}

\def\pnukpnumu{\(\Mpnukpnumu\)}
\def\pnukppippio{\(\Mpnukppippio\)}

\begin{document}

\baselineskip 14pt

\title{Current Status of Nucleon Decay Searches with Super-Kamiokande}
\author{Brett Viren for the Super-Kamiokande Collaboration}
\address{State University of New York at Stony Brook}
\maketitle

\begin{abstract}
  Evidence for Nucleon Decay has yet to be observed.  Current results
  from the observation of a 45 kton-year exposure of \superk{} and
  lifetime limits for nucleons to decay via lepton + pion, lepton +
  eta and lepton + kaon modes are presented.
\end{abstract}

\section{Motivation}

In the Standard Model it is assumed that the proton is completely
stable because there are no lighter products to which the proton could
decay without violating baryon number conservation.  However, the
requirement for an interaction to conserve baryon number is not backed
up by any symmetry in the Standard Model.  If baryon number
conservation is an incorrect assumption, then there are avenues for
protons (as well as neutrons which are in stable nuclei) to decay.

Furthermore, most theories which go beyond the Standard Model allow or
in most cases require, protons to be unstable at some level.  One of
the constraints on such theories is to predict a proton lifetime which
does not contradict current measured lifetime limits.  Experimental
limits on the proton lifetime can kill or constrain such theories
beyond the Standard Model.  Furthermore, the actuall observation of
proton decay would help open the door to new physics as well as give
answers to the ultimate state of matter in our universe in the far and
distant future.

One of the primary goals of the \superk{} experiment is to search for
proton (and neutron) decay.  The results of such searches performed
thus far are presented.

\section{\superk{} detector}

\superk{} is a ring imaging water \cherenkov{} detector containing 50 ktons
of ultra pure water held in a cylindrical stainless steel tank 1 km
underground in a mine in the Japanese alps.  The sensitive volume of
water is split into two parts.  The 2 m thick outer detector is viewed
with 1885 20 cm diameter photomultiplier tubes (PMTs) and acts
as a veto shield to tag incoming cosmic ray muons.  It completely
surrounds and is optically separated from the 33 m diameter, 36 m high
inner detector which is the primary sensitive volume.  The inner
detector contains 32.5 ktons of water and is viewed by 11146
inward pointing 50 cm diameter PMTs, giving a 40\% photocathode
coverage.

When relativistic charged particles pass through the water they emit
\cherenkov{} light at an angle of about \(42^\circ\) from the particles
direction of travel.  When this cone of light intersects with the
detector wall it is possible to image it as a ring.  By measuring the
charge produced in each PMT and the time at which it is collected, it
is possible to reconstruct the position and energy of the event as
well as the number, identity and momenta of the individual charged
particles in the event.

\section{Atmospheric Neutrino Background}

Many very exciting results have come from studying what 
is thought to be merely the
background to proton decay events.  Of particular note is the
continued confirmation~\cite{subgev,multigev} of the atmospheric
neutrino problem as well as finding evidence of massive
neutrinos~\cite{evidence} as the only conceivable solution.

For proton decay searches there are three classes of atmospheric
neutrino background events.  The first is that of inelastic charged
current events, \(\nu{}N \rightarrow N'\{e,\mu\}+n\pi{}\), where a
neutrino interacts with a nucleon in the water and produces a visible
lepton and some number of pions.  This can mimic proton decay modes
such as \peppo{}.\footnote{Thesis students: M. Shiozawa, B. Viren} The
second class is neutral current pion production, \(\nu{}N \rightarrow
\nu{}N'+n\pi{}\), the only visible products of which are pions.  This
is a background to, for example, \nnueta{}.\footnote{Thesis student:
  J.  Kameda} Finally, there is the mainstay of the atmospheric
neutrino group, single ring quasi elastic charged current, \(\nu{}N
\rightarrow N'\{\mu,e\}\), events which can look like, for example,
\pnukp{} decays.\footnote{Thesis students: M. Earl, Y. Hayato}

\section{The \peppo{} Mode}

The first mode discussed is \peppo{}.  Since this is one of the
simplest modes it serves well as a general example of proton decay
searches with \superk{} and will be discussed in some detail.

\begin{figure}[ht]
  \centerline{\epsfxsize=2in \epsfbox{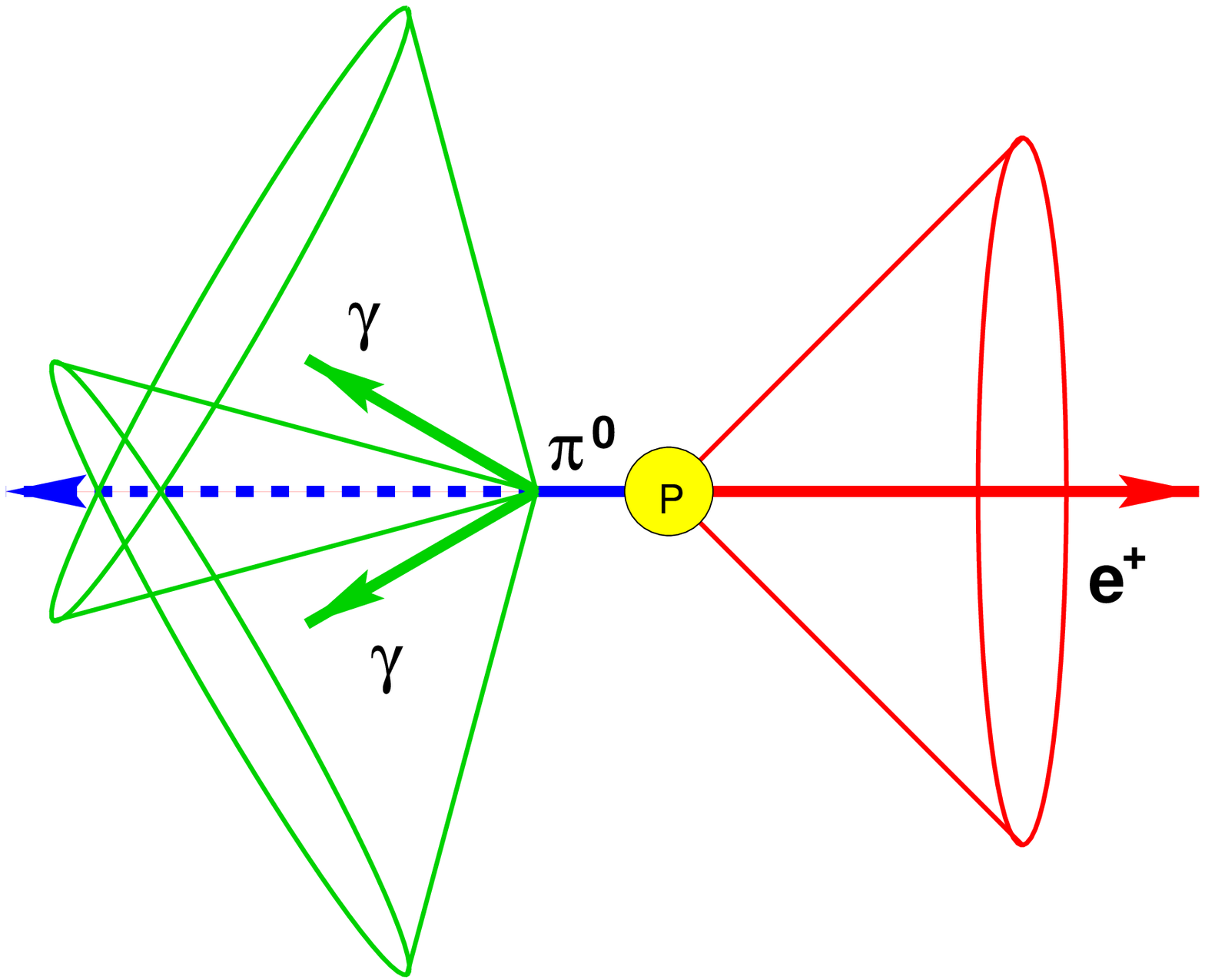}}
  \caption{Idealized \peppo{} decay in \superk{}.}
  \label{fig:ideal-peppo}
\end{figure}

Figure~\ref{fig:ideal-peppo} shows a cartoon of an ideal \peppo{}
decay.  Here, the positron, \(e^+\) and neutral pion, \(\pi^0\), exit
the decay region in opposite directions.  The positron initiates an
electromagnetic shower leading to a single isolated ring.  The
\(\pi^0\) will almost immediately decay to two photons which will go
on to initiate showers creating two, usually overlapping, rings.

\begin{figure}[htb]
  \centerline{\epsfxsize=5in \epsfbox{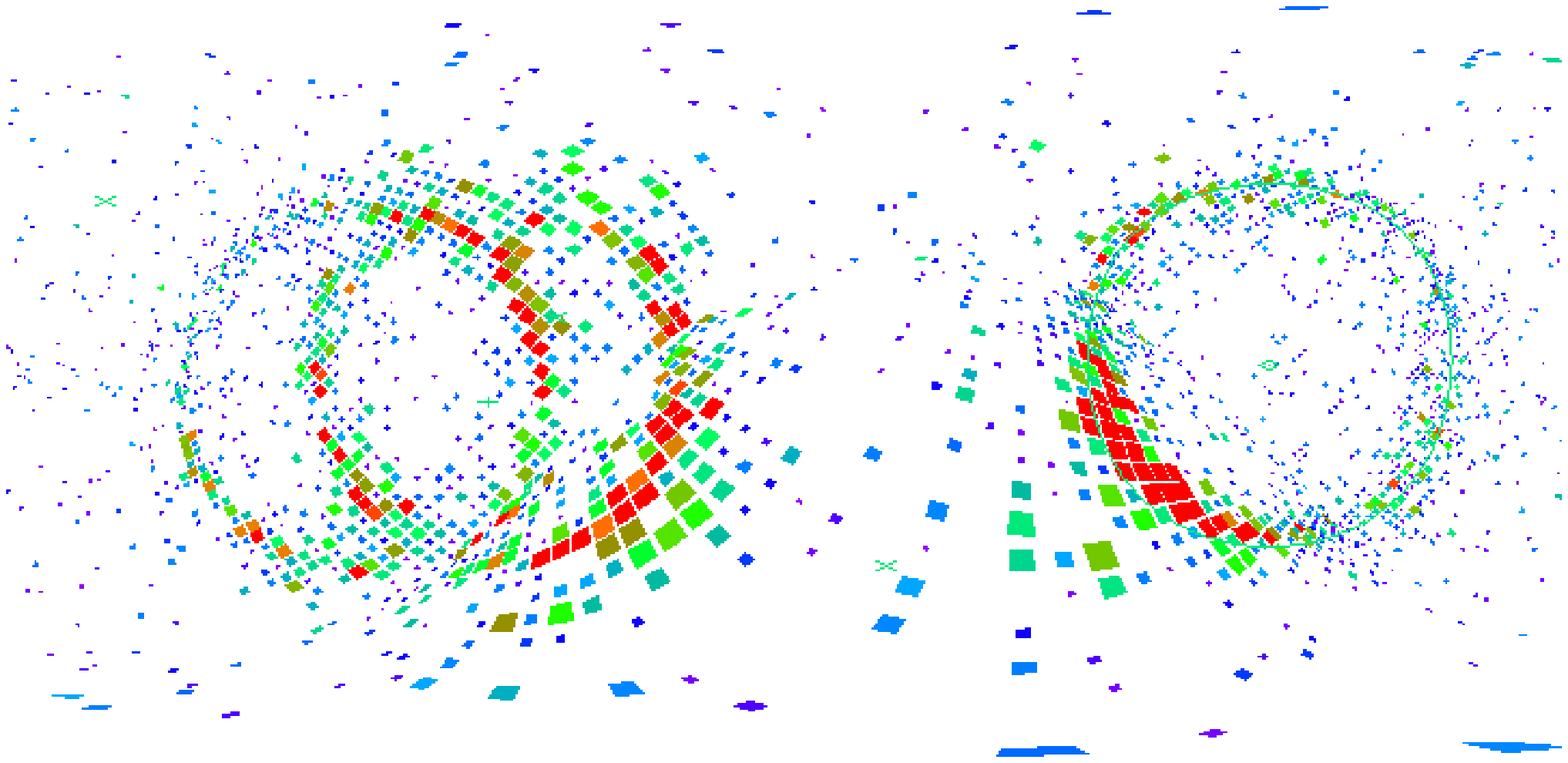}}
  \caption{\peppo{} MC event display.}
  \label{fig:peppomc}
\end{figure}

In \superk{}, such an ideal event might look like
Fig.~\ref{fig:peppomc}.  This event was generated with a detailed
\peppo{} event and detector Monte Carlo (MC) simulation. In this
figure, the PMTs are plotted as a function of \(\cos\theta\) {\it vs.}
\(\phi\) as viewed from the event vertex and are represented by
squares, colored by amount of collected charge (red is more, blue is
less) and sized to show distance from the event vertex.  The fuzzy
outer edges of the rings indicate an electromagnetic showering type of
ring.  Had the positron been replaced with a muon the single isolated
ring would have a sharp, distinct edge.

In general, real \peppo{} events will differ from this ideal picture
above because the pion can scatter or be absorbed entirely before it
exits the nucleus.  In addition the nuclear proton can have some
momentum due to Fermi motion.  These two effects, pion-nucleon
interaction and Fermi motion, result in a breaking of the balance of
reconstructed momentum.  In addition, the pion can decay
asymmetrically where one photon takes more than half of the pion's
energy leaving the second photon to create a faint or even completely
invisible ring.  These effects, plus energy resolution and systematic
uncertainties are considered when choosing the cuts to isolate
possible decay events from their background.

The same reduction of the 5 Hz raw (high energy) trigger rate at
\superk{} to the so called ``contained event sample'' (for more
information see~\cite{subgev,peppo97}) is used to find candidates for
proton decay events as well as atmospheric neutrino events.  From this
reduced data sample, selection criteria unique to each search are
applied to reduce the atmospheric background while keeping the
efficiency to find a particular decay mode high.

For the \peppo{} mode, the selection criteria are as follows: 
(A)~6000 \(< Q_{tot} <\) 9500 photoelectrons (PEs), 
(B)~2 or 3 e-like (showering type) rings, 
(C)~if 3 rings: 85 \(< M_{inv,\pi^0} <\) 185 MeV/c\(^2\), 
(D)~no decay electrons, 
(E)~800 MeV/c\(^2\) \(< M_{inv,tot} <\) 1050 MeV/c\(^2\), and
(F)~\(P_{tot} = \left|\sum\vec{P}_i\right|<\) 250 MeV/c.
The criterion (A) corresponds to a loose energy cut which reduces the
background with out much computation needed.  As stated above, it is
possible for one of the photons to be invisible, for which (B) allows.
If there are 3 rings criterion (C) requires that 2 of the rings
reconstruct to give a \(\pi^0\) mass.  Since there are no muons nor
charged pions expected, no decay electrons should be found and any
events which have them will be cut by (D).  Finally (E) requires the
total invariant mass to be near that of the proton and (F) requires
the total reconstructed momentum (magnitude of the vector sum of all
individual momenta) to be below the Fermi momentum for \(^{16}\)O.

\begin{figure}[htb]
  \centering
  \begin{minipage}[t]{2in}
    \epsfxsize=2in \epsfbox{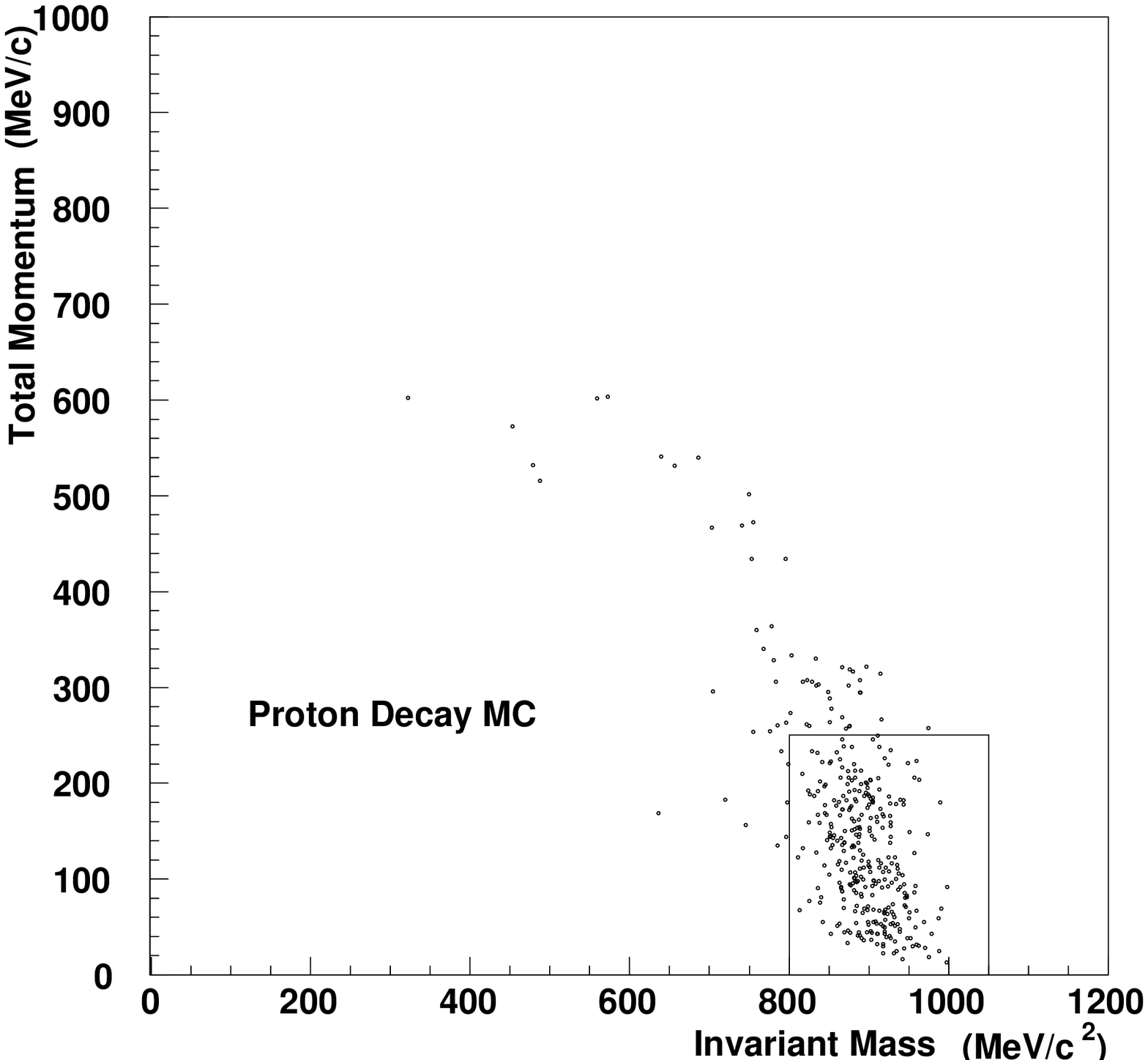}
  \end{minipage}
  \begin{minipage}[t]{2in}
    \epsfxsize=2in \epsfbox{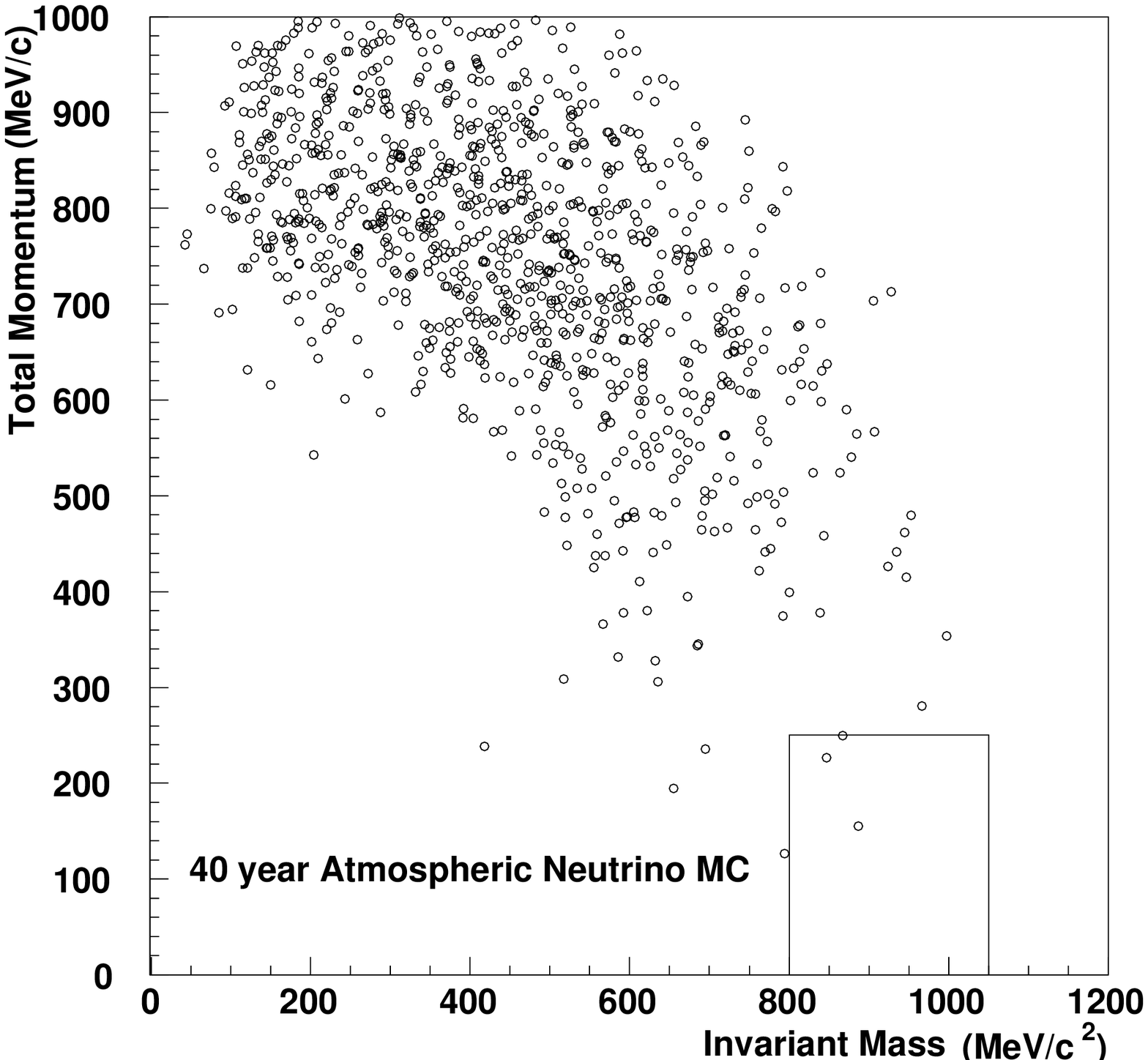}
  \end{minipage}
  \begin{minipage}[t]{2in}
    \epsfxsize=2in \epsfbox{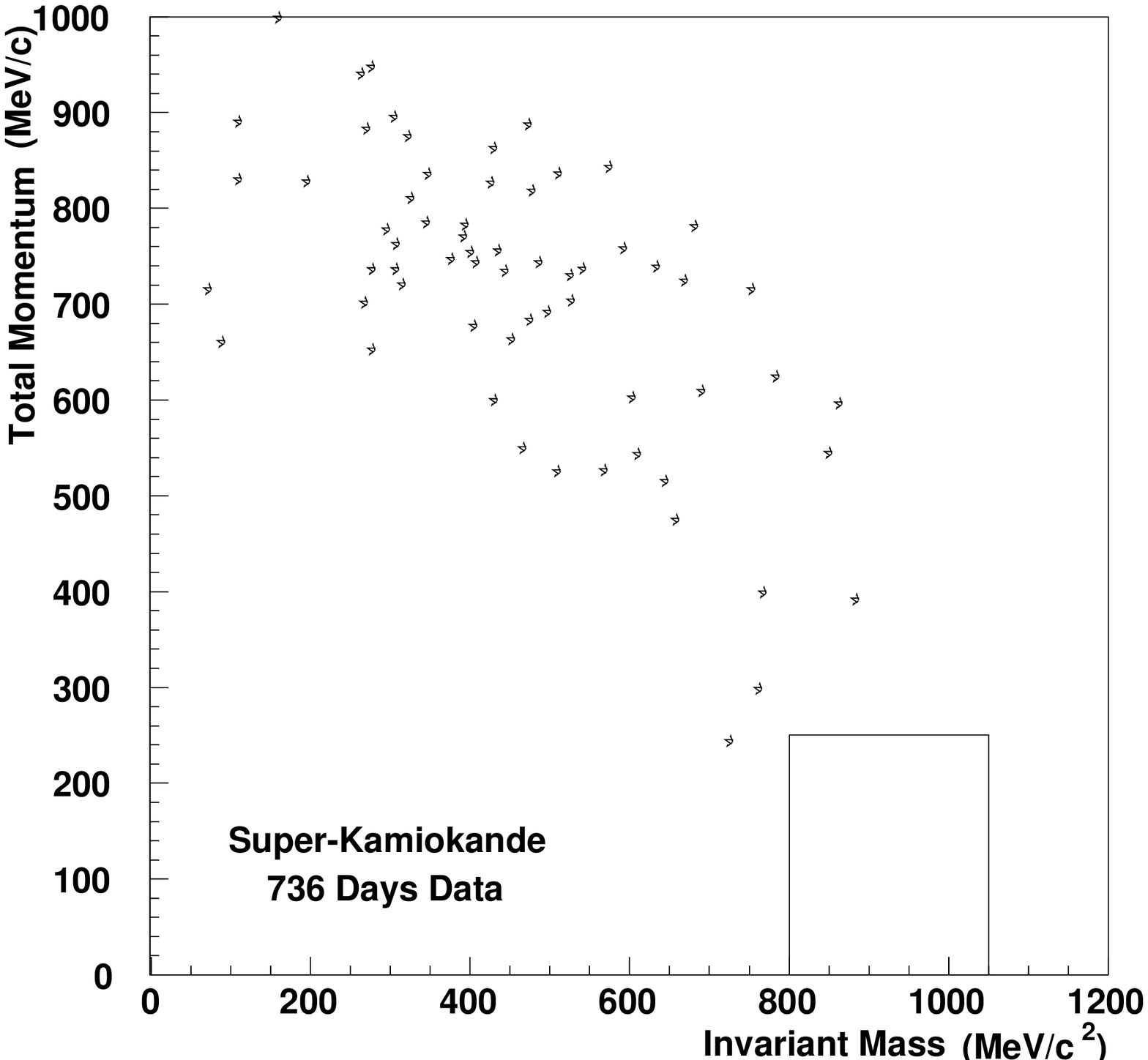}
  \end{minipage}
  \caption{\peppo{} mode.  Distributions of events in total reconstructed momentum {\it vs.} total invariant mass for (a) proton decay MC, (b) atmospheric neutrino MC, and (c) data.}
  \label{fig:peppo-massmom}
\end{figure}

Figure~\ref{fig:peppo-massmom} shows distributions of \peppo{} MC,
atmospheric MC, and data in reconstructed momentum {\it vs.}  invariant mass
after criteria (A)-(D) have been applied.  Criteria (E) and (F) shown
by the box.

When these criteria are applied to 45 kton\(\cdot\)years (736 days) 
of data we
find no candidate events.  Using atmospheric neutrino background MC
equivalent to 40 years of data taking it is estimated that 0.2
background events are expected in the data.  From MC simulations of
\peppo{} events, the efficiency to select any \peppo{} events in the
data sample is 44\%.  This gives a limit on the proton lifetime
divided by \peppo{} branching ratio (partial limit) of
\cllimit{\Mpeppo}{2.9}{33}.

\section{The \pmppo{} Mode}

The \pmppo{} mode is very similar to the \peppo{} mode.  The only
difference is to replace (A), (B) and (D) above with:
(A)~5000 \(< Q_{tot} <\) 7800 PE, 
(B)~1 \(\mu\)-like (non-showering) and 1 or 2 e-like rings, and 
(D)~1 decay electron,
respectively.  The region defined by criterion (A) is lower than in
the above case because a muon is more massive than a positron and
will go below \cherenkov{} threshold sooner, thus emitting less light.
The other two differences are also due to having a muon in the final
state instead of a positron.

When these criteria are applied, no decay candidates are found in the
data.  It is estimated 0.1 background events should exist and a
selection efficiency of 35\% is obtained.  The resulting partial limit
is \cllimit{\Mpmppo}{2.3}{33}.

\section{The \(\eta\) Modes}

Variations on the above two modes are found by replacing the neutral
pion by an eta particle decaying to two \(\gamma\)s.  This gives the
modes, \pepeta{}, \pmpeta{}, and \nnueta{}.

All selection criteria are then very similar to the two previous modes
except for (B) which is tightened up to only allow three ring events
(except for \nnueta{} which requires exactly two) and (D) which
requires the eta invariant mass to be reconstructed in the region
\(470 < M_{inv,\eta} < 610\) MeV/c\(^2\).

The \pepeta{} search results in no candidate events on top of an
expected background of 0.3 events, a selection efficiency of 17\% and
a partial lifetime limit of \cllimit{\Mpepeta ;\; \eta \rightarrow
  \gamma\gamma}{1.1}{33}.

For \pmpeta{}, no background is expected and no candidate events are
found.  The selection efficiency is 12\% and the partial lifetime
limit is \cllimit {\Mpmpeta ;\; \eta \rightarrow
  \gamma\gamma}{0.78}{33}

Finally, the \nnueta{} search finds 5 candidates consistent with the 9
expected background events and a selection efficiency of 21\%.  The
partial limit for this mode is \cllimit {\Mnnueta ;\; \eta \rightarrow
  \gamma\gamma}{0.56}{33}.

\section{The \pnukp{} Modes}

\begin{figure}[htb]
  \centering
  \begin{minipage}[t]{2in}
    \epsfxsize=2in \epsfbox{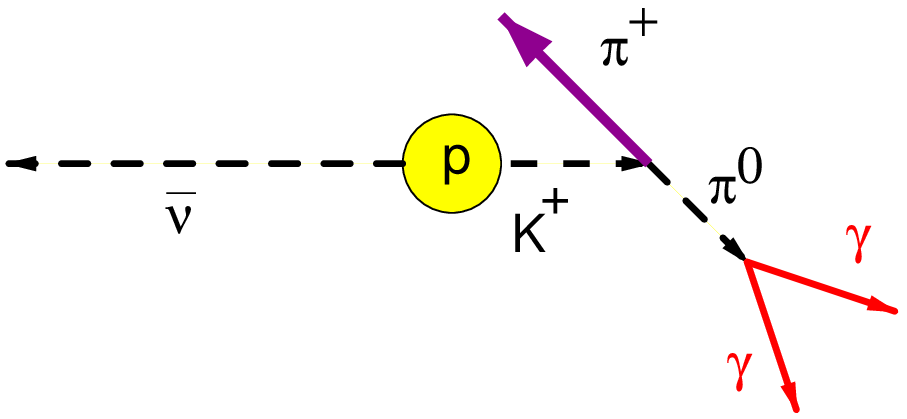}
  \end{minipage}
  \hspace{1in}
  \begin{minipage}[t]{2in}
    \epsfxsize=2in \epsfbox{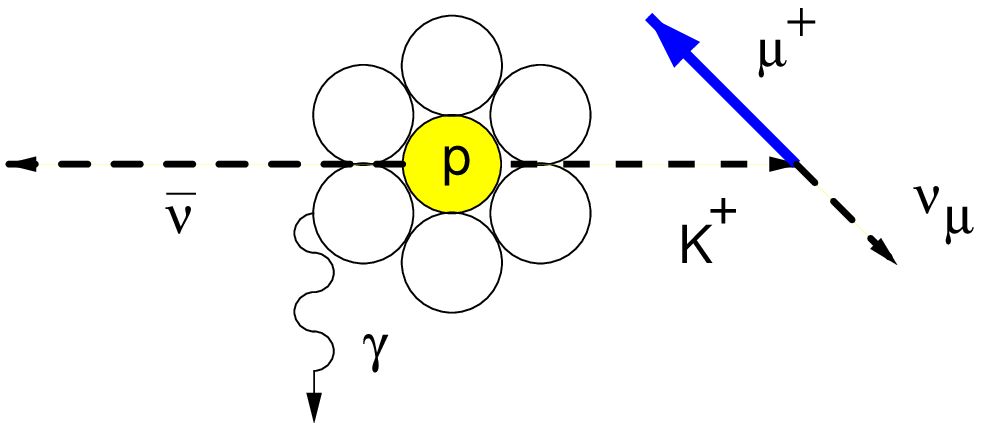}
  \end{minipage}
  \caption{\pnukp mode, \(K^+ \rightarrow{} \pi^+\pi^0\) and \(K^+ \rightarrow{} \mu^+\nu_\mu\) branches.}
  \label{fig:pnukp-cartoon}
\end{figure}

\superk{} searches for the \pnukp{} mode~\cite{pnukp-pdk} 
by looking for the products
from the two primary branches of the \(K^+\) decay.  These are
pictured in Fig.~\ref{fig:pnukp-cartoon}.  In the \(K^+ \rightarrow
\mu^+\nu_\mu\) case, when the decaying proton is in the \(^{16}\)O,
the nucleus will be left as an excited \(^{15}\)N.  Upon
de-excitation, a prompt 6.3 MeV photon will be emitted.  So this
second branch has two independent searches: one in which the signature
of this prompt photon is required and one in which it is explicitly
absent.  Unlike the other modes presented, the \pnukp{} search has
been done with data from only 33 kton\(\cdot\)years (535 days)
of exposure.

The criteria for the \pnukppippio{} search are as follows:
(A)~2 e-like rings,
(B)~1 decay electron,
(C)~\(85 < M_{inv,\pi^0} < 185\) MeV/c\(^2\),
(D)~\(175 < P_{\pi^0} < 250\) MeV/c,
(E)~\(40 < Q_{\pi^+} < 100\) PE.
The \(\pi^+\) is very close to \cherenkov{} threshold and is expected
to only produce a small amount of light as in (E).  Since this is not
enough to produce an identifiable ring only the rings from the 2
photons from the decay of the \(\pi^0\) are required in (A).  These
photons must reconstruct to an invariant mass in the range defined by
(C) as well as a momentum range defined in (D).

\begin{figure}[htb]
  \centering
  \begin{minipage}[t]{2in}
    \epsfxsize=2in \epsfbox{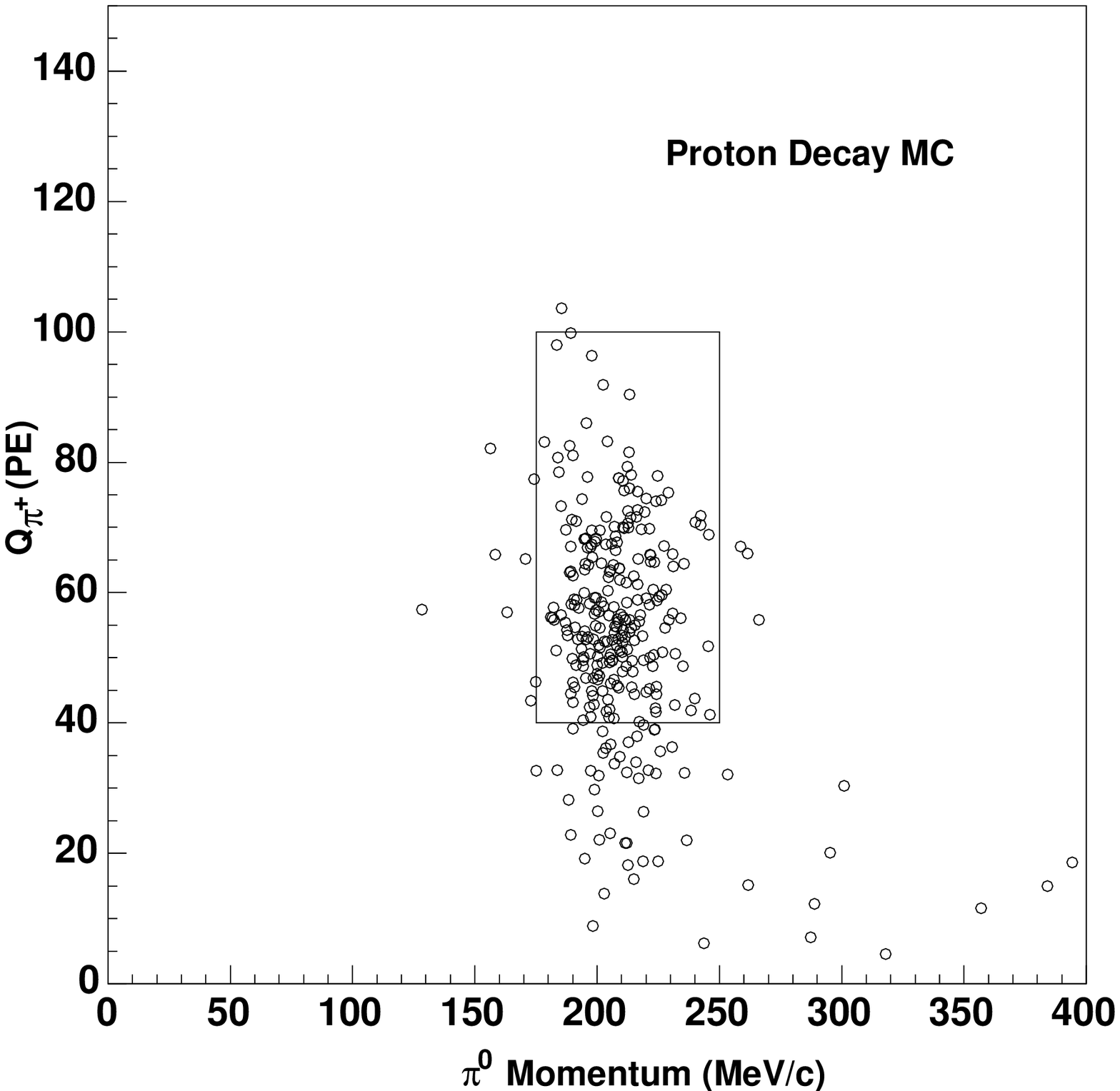}
  \end{minipage}
  \begin{minipage}[t]{2in}
    \epsfxsize=2in \epsfbox{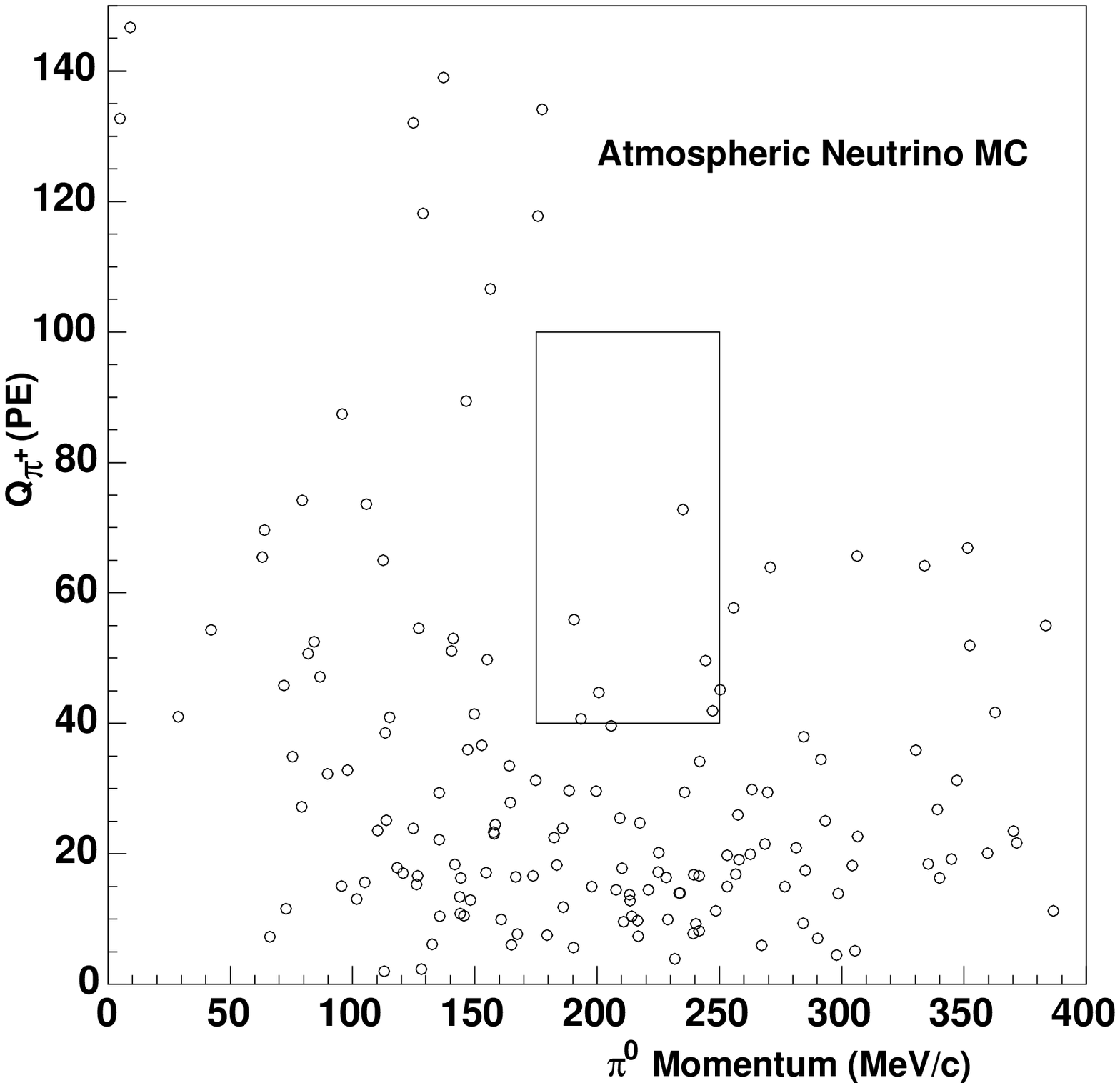}
  \end{minipage}
  \begin{minipage}[t]{2in}
    \epsfxsize=2in \epsfbox{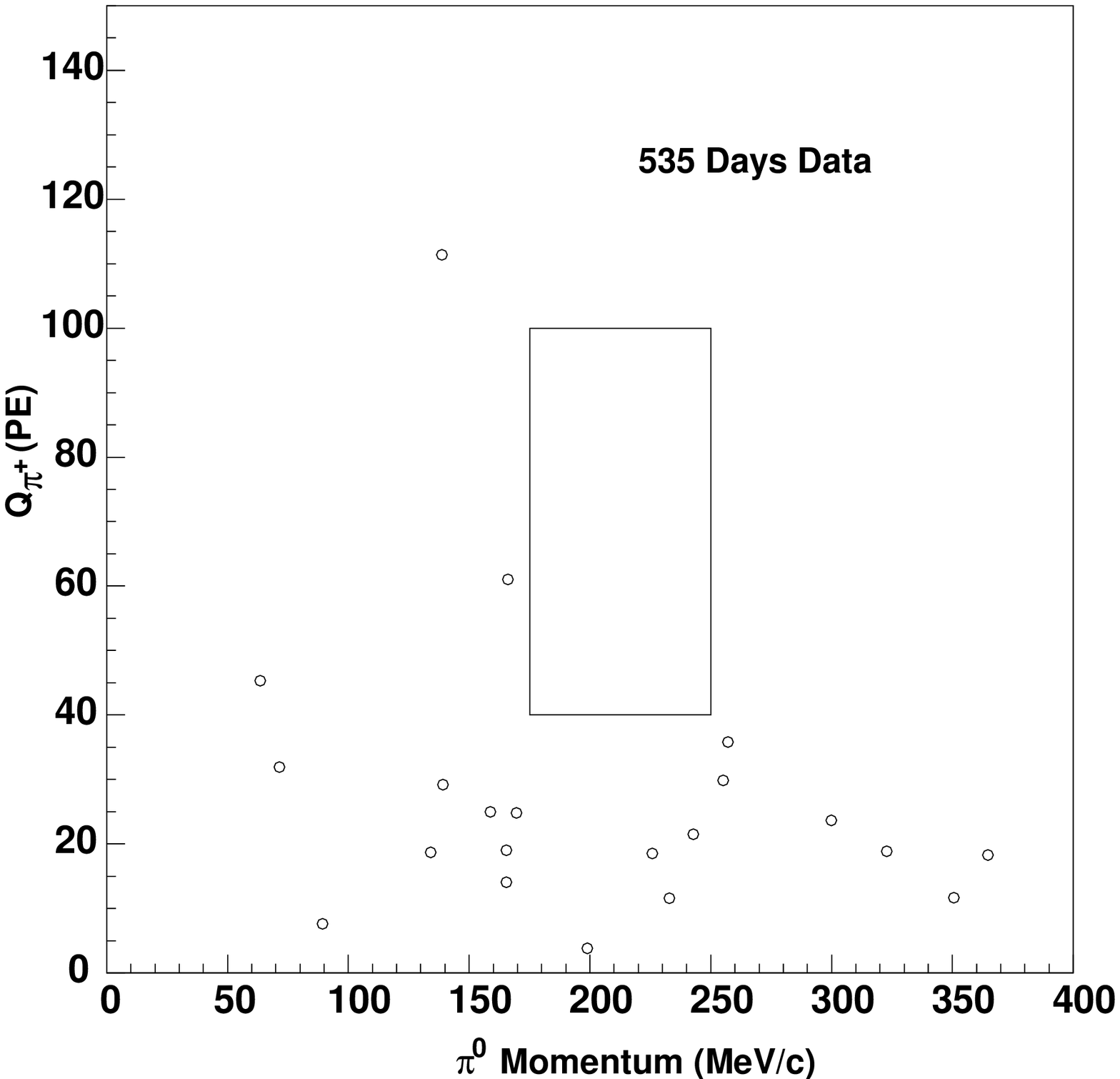}
  \end{minipage}
  \caption{\pnukppippio{} mode.  Distributions of events in \(Q_{\pi^+}\)  {\it vs.} \(\pi^0\) momentum for (a) proton decay MC, (b) atmospheric neutrino MC, and (c) data.}
  \label{fig:pnukppippio}
\end{figure}

Figure~\ref{fig:pnukppippio} shows the event distributions in
\(Q_{\pi^+}\) {\it vs.} \(\pi^0\) momentum for proton decay MC, atmospheric
neutrino MC and data after criteria (A) through (C).  Criteria (D) and
(E) are represented by the box.

No candidates are found and 0.7 background events are expected.  The
selection efficiency is 6.5\%, giving a partial lifetime limit of
\cllimit{\Mpnukppippio}{3.1}{32}.

When searching for the \pnukpnumu{} with a 6.3 MeV prompt photon
search the following criteria are required:
(A)~1 \(\mu\)-like ring,
(B)~1 decay electron,
(C)~\(215 < P_{\mu} < 260\) MeV/c, and
(D)~\(N_{PMT}>7 ; 12 < t_{PMT} < 120\) ns before \(\mu\) signal.
The only particle giving a visible ring is the mono-energetic muon.
Criteria (A-C) select for that.  In criterion (D), \(t_{PMT}\) is the
time a PMT was hit subtracted by the time it would take a photon to
get directly from the fit event vertex to the PMT (so called ``time
minus time of flight'' or ``timing residual'').  This requirement to
have a significant number of hit PMTs within 1 to 10 kaon lifetimes is
illustrated in Fig.~\ref{fig:pnuk-gamma}.

\begin{center}
\begin{minipage}[b]{3in}
  \begin{figure}[htb]
    \epsfxsize=3in \epsfbox{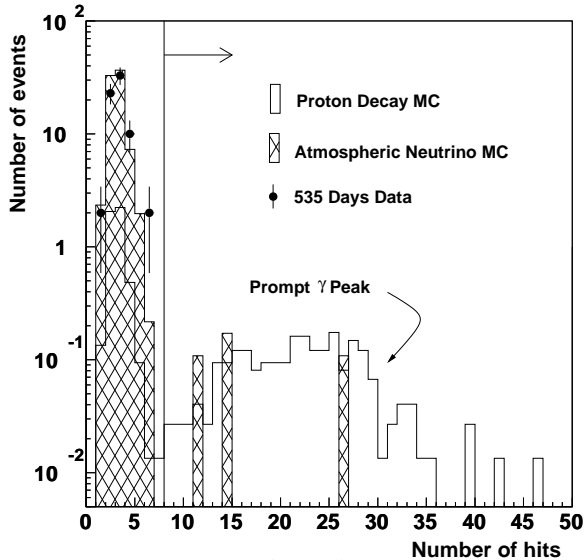}
    \caption{\pnukpnumu{} mode with prompt \(\gamma\) signature.  The atmospheric neutrino MC (hatched histogram) and proton decay MC (empty histogram) has been normalize to the data (dots).}
    \label{fig:pnuk-gamma}
  \end{figure}
\end{minipage}\hspace{0.5in}
\begin{minipage}[b]{3in}
  \begin{figure}[htb]
    \epsfxsize=3in \epsfbox{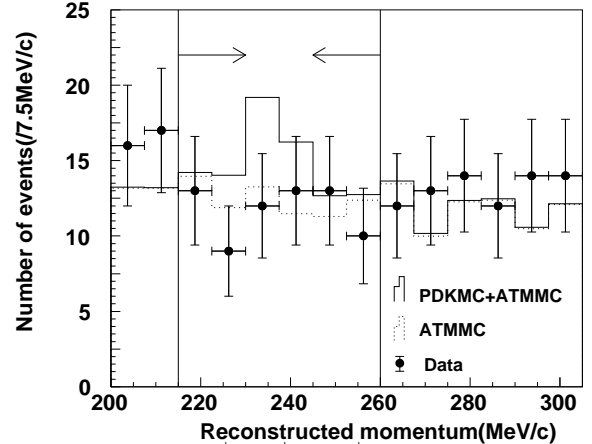}
    \caption{\pnukpnumu{} with out prompt \(\gamma\) signature.  The solid PDKMC+ATMMC histogram shows the atmospheric neutrino MC plus the proton decay MC assuming a proton lifetime that of the found 90\% CL.  The dotted ATMMC histogram is just the atmospheric neutrino MC normalized to the data which are the dots.}
    \label{fig:pnuk-nogamma}
  \end{figure}
\end{minipage}
\end{center}

This search has an efficiency of 4.4\%, finds no candidates on an
estimated background of 0.4 events, and sets a limit of
\cllimit{\Mpnukpnumug}{2.1}{32}.

The complimentary case where no prompt gamma is allowed has the same
criteria as the prompt gamma case except for the last:
(D)~\(N_{PMT}<=7 ; 12 < t_{PMT} < 120\) ns before \(\mu\) signal.
Since this allows a significant amount of background to survive the
selection criteria the limit is set by fitting for an excess of proton
decay events above the atmospheric neutrino background in the
reconstructed momentum spectrum.  This is shown in
Fig.~\ref{fig:pnuk-nogamma}.  What is found is that, if anything, the
data fluctuates downward in the region of expected muon momentum.

In this region, 70 candidate events are found which is consistent with
the 74.5 events expected from atmospheric neutrinos.  With an
efficiency of 40\% a limit of \cllimit{\Mpnukpnumunog}{3.3}{32} is
found.

The combined limit for these three topologies is
\cllimit{\Mpnukpnumu}{6.8}{32}.

\section{The \pmpko{} Mode}

\begin{figure}[htb]
  \centering
  \begin{minipage}[t]{2in}
    \epsfxsize=2in \epsfbox{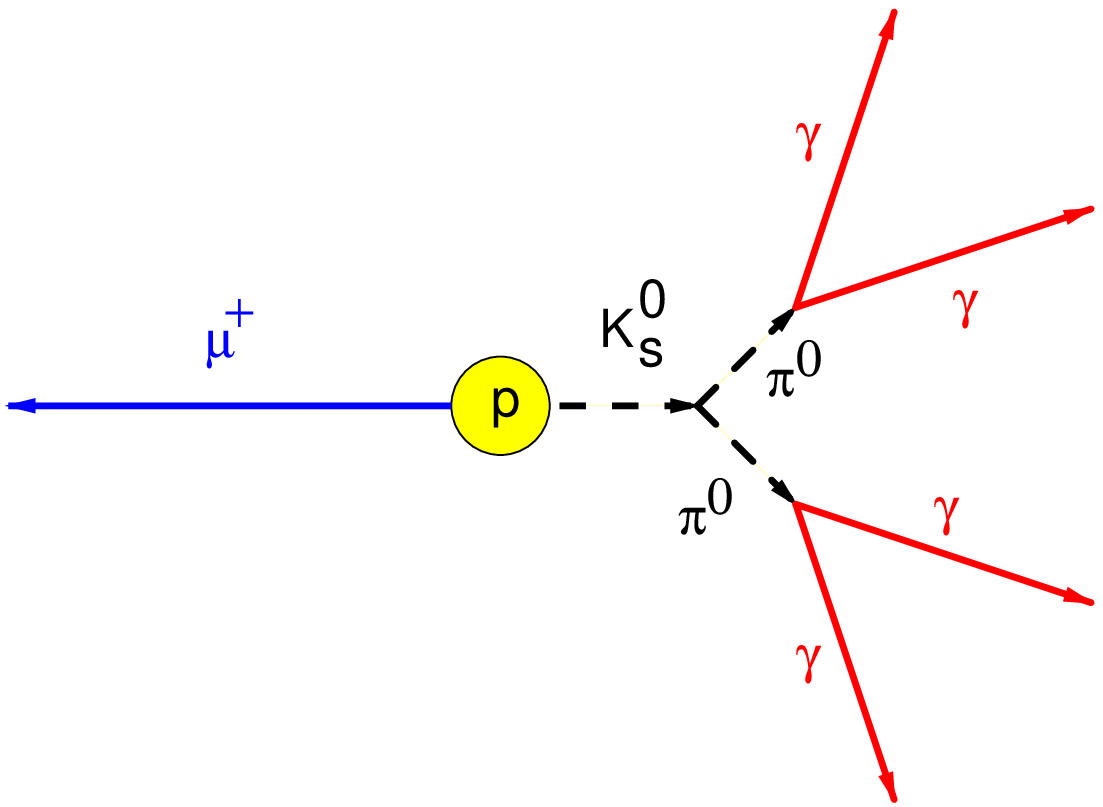}
  \end{minipage}
  \caption{\pmpko{} mode.}
  \label{fig:pmuko-cartoon}
\end{figure}

The search for proton decay via \pmpko{} so far relies on \(K^0_s
\rightarrow \pi^0\pi^0 \rightarrow \gamma\gamma\gamma\gamma\) decay
branch of the kaon.  The selection criteria for this are:
(A)~1 \(\mu\)-like and 2 or more e-like rings,
(B)~1 or less decay electron,
(C)~\(400 < M_{inv,K^0} < 600\) MeV/c\(^2\),
(D)~\(150 < P_\mu < 400\) MeV/c,
(E)~\(750 < M_{inv,tot} < 1000\) MeV/c\(^2\), and
(F)~\(P_{tot} < 300\) MeV/c.
Because it is possible for one photon from each pion to be missed,
only 2 are required in (A) in order to preserve a high efficiency.
Using all e-like rings, a reconstructed \(K^0\) mass is required in
(C).  The reconstructed muon momentum should be a discrete value
smeared by Fermi motion and energy resolution as is required in (D).
Finally, (E) and (F) require total invariant mass near proton rest
mass and total reconstructed momentum to be less than the Fermi
momentum.

\begin{figure}[htb]
  \centering
  \begin{minipage}[t]{3in}
    \epsfxsize=3in \epsfbox{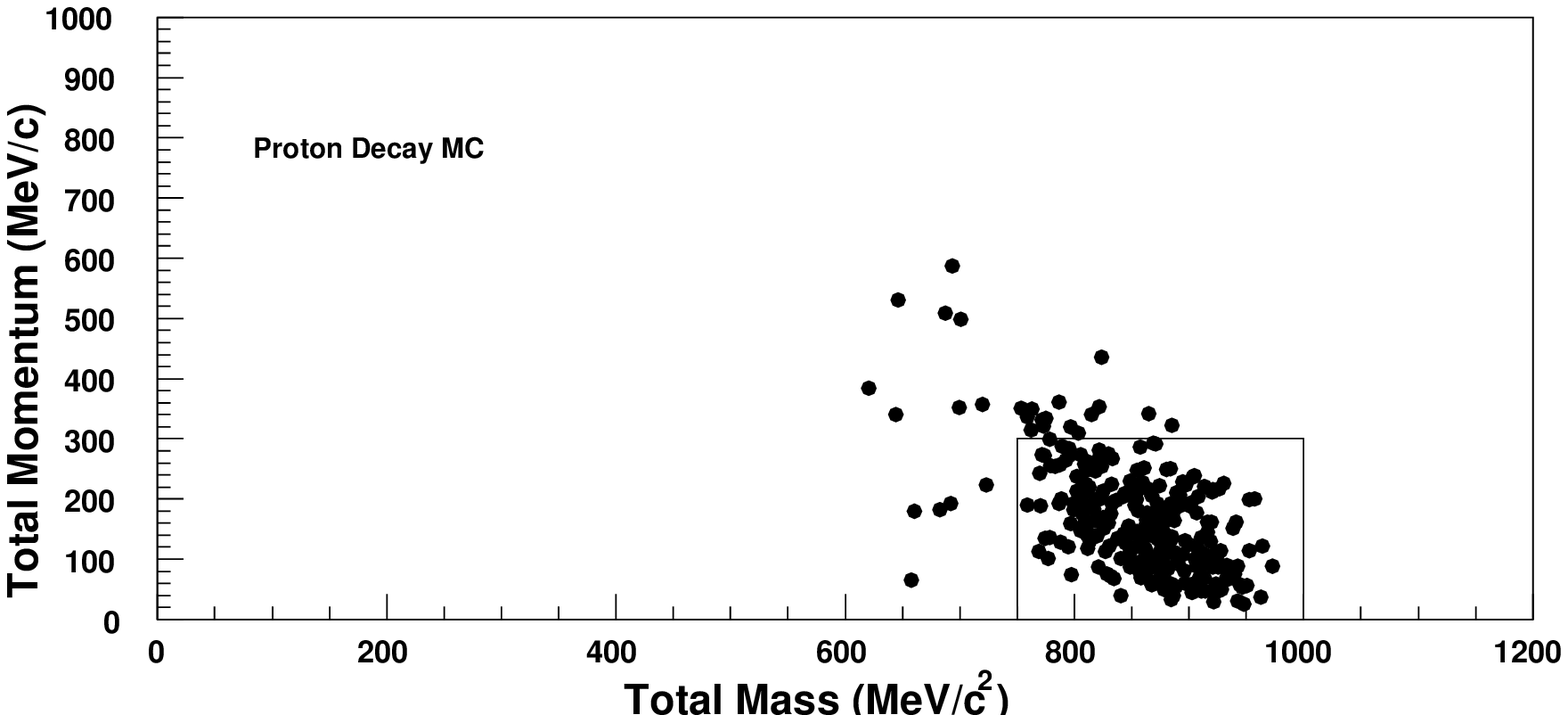}
  \end{minipage}
  \begin{minipage}[t]{3in}
    \epsfxsize=3in \epsfbox{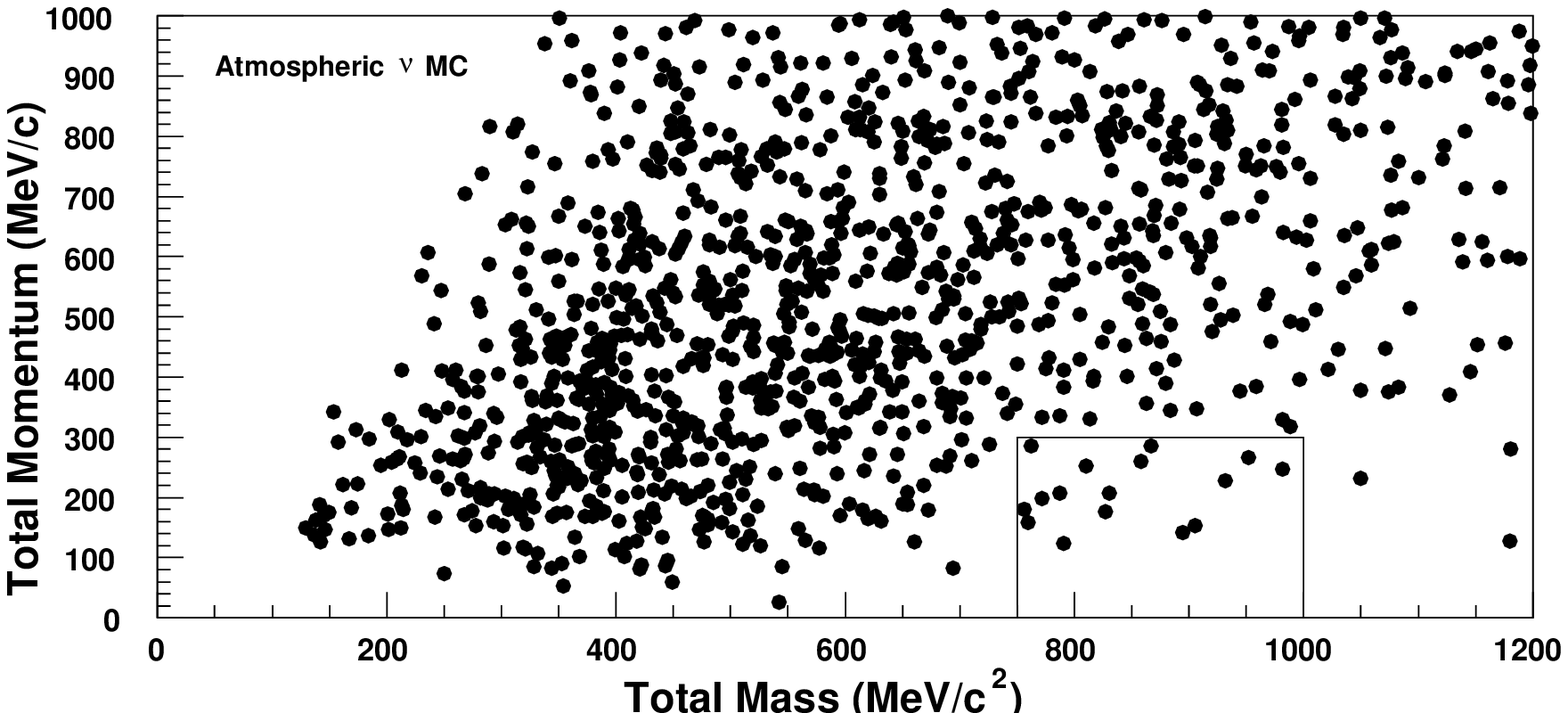}
  \end{minipage}
  \begin{minipage}[t]{3in}
    \vspace{1mm}
    \epsfxsize=3in \epsfbox{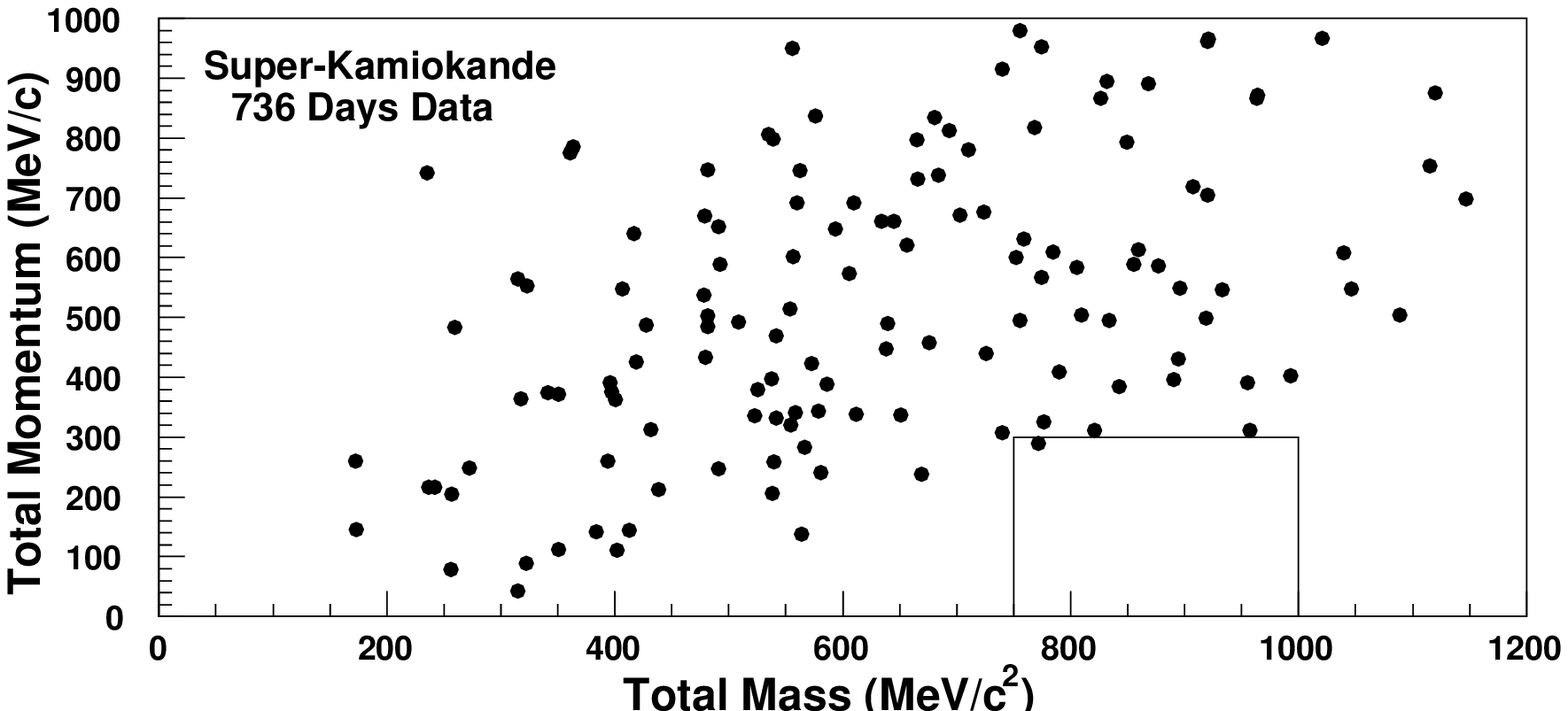}
  \end{minipage}
  \caption{\pmpko{} mode. Distributions of events in total reconstructed momentum {\it vs.} total invariant mass for (a) proton decay MC, (b) atmospheric \(\nu\) MC and (c) data.}
  \label{fig:pmuko-results}
\end{figure}

Figure~\ref{fig:pmuko-results} shows the event
distributions in total momentum {\it vs.} total invariant mass for proton
decay MC, atmospheric neutrino MC and the data for events which pass
criteria (A) and (B).  The boxes represent criteria (E) and (F).  The
single data event which is inside the box is far outside the
requirements of (C) and (D) and so is not a valid candidate.

This search was done on 45 kton\(\cdot\)years of data and has a 6.1\%
efficiency.  No candidates were found and 0.65 background events were
expected.  The resulting limit is \cllimit{\Mpmpko}{4.0}{32}.

\section{Summary of searches}

The current \superk{} nucleon decay results are summarized in
Table~\ref{table:summary}.  As a comparison, limits collected in PDG
1998~\cite{pdg} are also included.  Finally, although the best
limits have been set by large water \cherenkov experiments, iron
calorimeters offer a complementary search with tracking sensitivity to
kaons and low momentum pions and muons.  Recent results from the
Soudan 2 experiment were also presented and are listed below.

\begin{table}
  \begin{center}
    \caption{Summary of nucleon decay searches.  All limits are in units of \(10^{32}\) years and are at 90\% confidence level.}
    \begin{minipage}{3.5in}
      \begin{tabular}{|c|c|c|c|}
        Mode & \superk{} & PDG 1998 & Soudan 2 \\
        \hline
        \peppo{}    & 29    & 5.5  & --    \\
        \hline
        \pmppo{}    & 23    & 2.7  & --    \\
        \hline
        \pepeta{}   & 11    & 1.4  & 0.70  \\
        \hline
        \pmpeta{}   & 7.8   & 0.69 & 0.78   \\
        \hline
        \nnueta{}   & 5.6   & 0.54 & 0.64   \\
        \hline
        \pnukp{}    & 6.8   & 1.0  & 0.78  \\
        \hline
        \pepko{}    & --    & 0.76 & 0.72   \\
        \hline
        \pmpko{}    & 4.0   & 0.64 & 1.04   \\
        \hline
        \(n \rightarrow \bar{\nu} K^0_s\) & -- & 0.86 & 0.62 
      \end{tabular}
    \end{minipage}
    \label{table:summary}
  \end{center}
\end{table}

\end{document}